\documentclass[nocrop]{bioinfo}

\usepackage{algorithm}
\usepackage{algorithmic}
\usepackage{amsmath}
\usepackage{amssymb} 
\usepackage{amsthm} 
\usepackage{siunitx}
\usepackage{booktabs}
\usepackage{float}
\usepackage{color}

\copyrightyear{2018} \pubyear{2018}
\newcommand{\red}[1]{\textcolor{black}{#1}}

\access{Advance Access Publication Date: Day Month Year}
\appnotes{Manuscript Category}

\begin{document}
\firstpage{1}
\setlength{\abovedisplayskip}{3pt} 
\setlength{\belowdisplayskip}{3pt} 

\subtitle{Structural Bioinformatics}
\title[DeepAffinity: Interpretable Deep Learning of Compound--Protein Affinity]{DeepAffinity: Interpretable Deep Learning of Compound-Protein Affinity through Unified Recurrent and Convolutional Neural Networks}
\author[Karimi, Wu, Wang and Shen]{Mostafa Karimi\,$^\text{\sfb 1,2}$, Di Wu\,$^\text{\sfb 1}$, Zhangyang Wang\,$^\text{\sfb 3}$ and Yang shen\,$^{\text{\sfb 1,2,*}}$}
\address{$^{\text{\sf 1}}$Department of Electrical and Computer Engineering, $^{\text{\sf 2}}$TEES--AgriLife Center for Bioinformatics and Genomic Systems Engineering, and $^{\text{\sf 3}}$Department of Computer Science and Engineering, Texas A\&M University, College Station, 77843, USA.}

\corresp{$^\ast$To whom correspondence should be addressed.}

\history{Received on XXXXX; revised on XXXXX; accepted on XXXXX}

\editor{Associate Editor: XXXXXXX}

\abstract{
\textbf{Motivation:} Drug discovery demands rapid quantification of compound-protein interaction  (CPI). 
However, there is a lack of methods that can predict compound-protein affinity from sequences alone with high applicability, accuracy, and interpretability.   
\\
\textbf{Results:}  We present a seamless integration of domain knowledges and learning-based approaches. Under novel representations of structurally-annotated protein sequences, a semi-supervised deep learning model that unifies recurrent and convolutional neural networks has been proposed to exploit both unlabeled and labeled data, for jointly encoding molecular representations and predicting affinities. Our representations and models outperform conventional options in achieving relative error in IC$_{50}$ within 5-fold for test cases and \red{20}-fold for protein classes not included for training. Performances for new protein classes with few labeled data are further improved by transfer learning.
Furthermore, \red{separate and joint} attention mechanism\red{s} \red{are developed and } embedded to our model to add to its interpretability, as illustrated in case studies for predicting and explaining selective drug-target interactions. \red{Lastly, alternative representations using protein sequences or compound graphs and  a unified RNN/GCNN-CNN model using graph CNN (GCNN) are also explored to reveal algorithmic challenges ahead.}\\
\textbf{Availability:} Data and source codes are available at https://github.com/Shen-Lab/DeepAffinity\\
\textbf{Contact:} \href{yshen@tamu.edu}{yshen@tamu.edu}\\
\textbf{Supplementary information:} Supplementary data are available at \\\href{http://shen-lab.github.io/deep-affinity-bioinf18-supp-rev.pdf}{http://shen-lab.github.io/deep-affinity-bioinf18-supp-rev.pdf}.
}

\maketitle
\section{Introduction}
Drugs are often developed to target proteins that participate in many cellular processes.  Among almost 900 FDA-approved drugs as of year 2016, over 80\% are small-molecule compounds that act on proteins for drug effects~\citep{Santos2017}. Clearly, it is of critical importance to characterize compound-protein interaction for drug discovery and development, whether screening compound libraries for given protein targets to achieve desired effects or testing given compounds against possible off-target proteins to avoid undesired effects. However, experimental characterization of every possible compound-protein pair can be daunting, if not impossible, considering the enormous chemical and proteomic spaces. Computational prediction of compound-protein interaction  (CPI) has therefore made much progress recently, especially for repurposing and repositioning known drugs for previously unknown but desired new targets~\citep{keiser2009predicting,PowerJAMA2014} and for anticipating compound side-effects or even toxicity due to interactions with off-targets or other drugs~\citep{Chang2010PLOS, DeepTox}. 

Structure-based methods can predict compound-protein affinity, i.e., how active or tight-binding a compound is to a protein; and their results are highly interpretable. This is enabled by evaluating energy models~\citep{gilson2007calculation} 
on 3D structures of protein-compound complexes. As these structures are often unavailable, they often need to be first predicted by  ``docking" individual structures of proteins and compounds together before their energies can be evaluated, which tends to be a bottleneck for computational speed and accuracy~\citep{Leach2006ProtLig}. Machine learning has been used to improve scoring accuracy based on energy features
~\citep{MLScoringRev2015}.

More recently, deep learning has been introduced to predict compound activity or binding-affinity from 3D structures directly. Wallach et al. developed AtomNet, a deep convolutional neural network  (CNN), for modeling bioactivity and chemical interactions~\citep{wallach2015atomnet}. 
Gomes et al.~\citep{gomes2017atomic} developed atomic convolutional neural network  (ACNN) for binding affinity by generating new pooling and convolutional layers specific to atoms. Jimenez et al.~\citep{Jimenez2018} also used 3D CNN with molecular representation of 3D voxels assigned to various physicochemical property channels. Besides these 3D CNN methods, Cang and Wei represented 3D structures in novel 1D topology invariants in multiple channels for CNN~\citep{TopologyNet17}.  These deep learning methods often improve scoring thanks to modeling long-range and multi-body atomic interactions.  Nevertheless, they still rely on actual 3D structures of CPI and remain largely untested on lower-quality structures predicted from docking, which prevents large-scale applications. 

Sequence-based methods overcome the limited availability of structural data and the costly need of molecular docking. Rather, they exploit rich omics-scale data of protein sequences, compound sequences  (e.g. 1D binary substructure fingerprints~\citep{wang2009pubchem}) and beyond  (e.g. biological networks).  However, they have been restricted to classifying CPIs~\citep{Chen2016CPIRev}  mainly into two types  (binding or not) and occasionally more  (e.g., binding, activating, or inhibiting~\citep{WangAndZeng2013}). And more importantly, their interpretablity is rather limited due to high-level features.  Earlier sequence-based machine learning methods are based on shallow models for supervised learning, such as support vector machines, logistic regression, random forest, and shallow neural networks ~\citep{cheng2012prediction_3, yu2012systematic,tabei2013scalable,shi2013protein, cheng2016effectively}. 
These shallow models are not lack of interpretability \textit{per se}, but the sequence-based high-level features do not provide enough interpretability for mechanistic insights on why a compound--protein pair interacts or not.  

Deep learning has been introduced to improve CPI identification from sequence data and shown to outperform shallow models.  Wang and Zeng developed a method to predict three types of CPI based on restricted Boltzmann machines, a two-layer probabilistic graphical model and a type of building block for deep neural networks~\citep{WangAndZeng2013}.  Tian et al. boosted the performance of traditional shallow-learning methods by a deep learning-based algorithm for CPI~\citep{tian2016boosting}. Wan et al. exploited feature embedding algorithm such as latent semantic algorithm ~\citep{deerwester1990indexing} and word2vec ~\citep{mikolov2013efficient} to automatically learn low-dimensional feature vectors of compounds and proteins from the corresponding large-scale unlabeled data~\citep{wan2016deep}. Later, they trained deep learning to predict the likelihood of their interaction by exploiting the learned low-dimensional feature space.  However, these deep-learning methods inherit from sequence-based methods two limitations: simplified task of predicting whether rather than how active CPIs occur as well as low interpretability due to the lack of fine-resolution structures. In addition, interpretability for deep learning models remains a challenge albeit with fast progress especially in a model-agnostic setting~\citep{LIME16,pmlr-v70-koh17a} .    

As has been reviewed, structure-based methods predict quantitative levels of CPI in a realistic setting and are highly interpretable with structural details. But their applicability is restricted by the availability of structure data, and the molecular docking step makes the bottleneck of their efficiency. Meanwhile, sequence-based methods often only predict binary outcomes of CPI in a simplified setting and are less interpretable in lack of mechanism-revealing features or representations; but they are broadly applicable with access to large-scale omics data and generally fast with no need of molecular docking.  

Our goal is to, realistically, predict quantitative levels of CPIs  (compound-protein affinity measured in IC$_{50}$, $K_i$, or $K_d$) from sequence data alone and to balance the trade-offs of previous structure- or sequence-based methods for broad applicability, high throughput and more interpretability. From the perspective of machine learning, this is a much more challenging regression problem compared to the classification problem seen in previous sequence-based methods. 

To tackle the problem, we have designed interpretable yet compact data representations and introduced a novel and interpretable deep learning framework that takes advantage of both unlabeled and labeled data. 
Specifically, we first have represented compound sequences in the Simplified Molecular-Input Line-Entry System  (SMILES) format~\citep{weininger1988smiles} and protein sequences in novel alphabets of structural and physicochemical properties. These representations are much lower-dimensional and more informative compared to previously-adopted small-molecule substructure fingerprints or protein Pfam domains ~\citep{tian2016boosting}. We then leverage the wealth of abundant unlabeled data to distill representations capturing long-term, nonlinear dependencies among residues/atoms in proteins/compounds, by pre-training bidirectional recurrent neural networks  (RNNs) as part of the seq2seq auto-encoder that finds much success in modeling sequence data in natural language processing~\citep{kalchbrenner2013recurrent}. And we develop a novel deep learning model unifying RNNs and convolutional neural networks  (CNNs), to be trained from end to end~\citep{wang2016studying} using labeled data for task-specific representations and predictions. Furthermore, we introduce \red{several} attention mechanisms to interpret predictions by isolating main contributors of molecular fragments \red{or their pairs}, which is further exploited for predicting binding sites and origins of binding specificity.  \red{Lastly, we explore alternative representations using  protein sequences or compound graphs  (structural formulae), develop graph CNN (GCNN) in our unified RNN/GCNN-CNN model, and discuss remaining challenges.}

\begin{figure*}[htb]
    \centering
     \vspace{-2em}
    \includegraphics[width=0.7\textwidth]{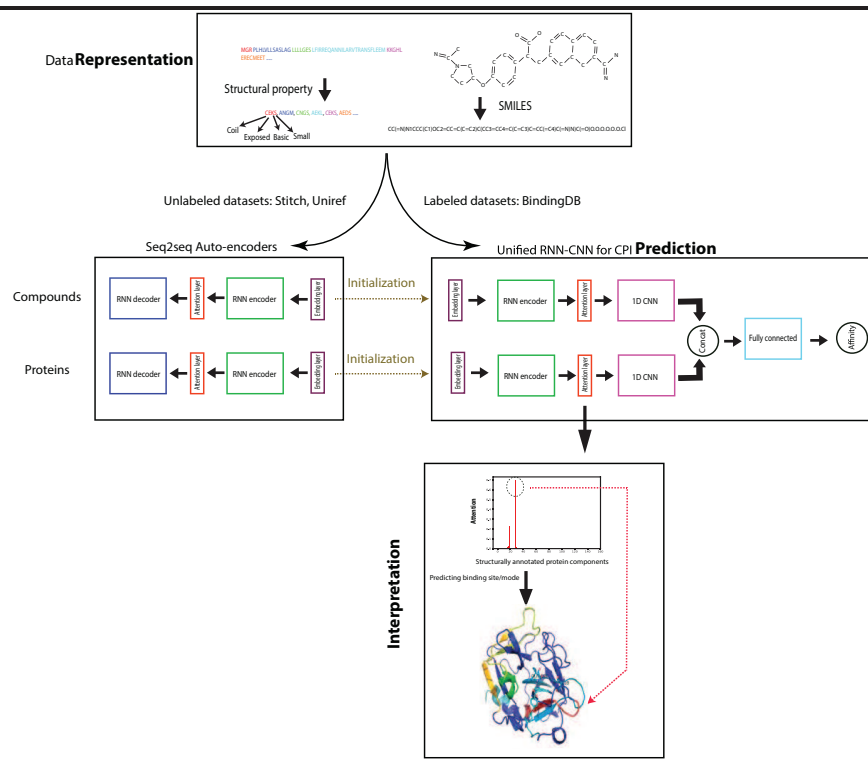}
        \vspace{-2em}
    \caption{The pipeline of our unified RNN-CNN method to predict \red{and interpret} compound-protein affinity.}
    \vspace{-1em}
    \label{fig:whole_CPI}
     \vspace{-1em}
\end{figure*}

The overall pipeline of our unified RNN-CNN method for semi-supervised learning  (data representation, unsupervised learning, and joint supervised learning) is illustrated in Fig.~\ref{fig:whole_CPI} with details given next.

\vspace{-3em}

\section{Materials and Methods}
\subsection{Data}

We used molecular data from three public datasets: labeled compound-protein binding data from BindingDB~\citep{liu2006bindingdb}, compound data in the SMILES format from STITCH~\citep{kuhn2007stitch} and protein amino-acid sequences from UniRef ~\citep{suzek2014uniref}.

From \red{489,280} IC$_{50}$-labeled samples collected from BindingDB, we completely excluded \red{four} classes of proteins from the training set:  nuclear estrogen receptors  (ER;  \red{3,374} samples), ion channels  (\red{14,599} samples), \red{ receptor tyrosine kinases  (34,318 samples)}, and G-protein-coupled receptors  (GPCR; \red{60,238} samples), to test the generalizability of our framework.  And we randomly split the rest into the training  (\red{263,583} samples including 10\% held out for validation) and the default test set  (\red{113,168} samples) without the aforementioned \red{four} classes of protein targets.  Similarly, we split a $K_i$  ($K_d$) labeled dataset into \red{101,134  (8,778)} samples for training, \red{43,391  (3,811)} for testing, \red{516  (4)} for ERs, \red{8,101  (366)} for ion channels, \red{3,355  (2,306) for tyrosine kinases}, and \red{77,994  (2,554)} for GPCRs. All labels are in logarithm forms: pIC$_{50}$, p$K_i$, and p$K_d$.  More details can be found in Sec. 1.1 of Supplementary Data.  
 
For unlabeled compound data from STITCH, we randomly chose 500K samples for training and 500K samples for validation  (sizes were restricted due to computing resources) and then removed those whose SMILES string lengths are above 100, resulting in 499,429 samples for training and 484,481 for validation. For unlabeled protein data from UniRef, we used all UniRef50 samples  (50\% sequence-identity level) less those of  lengths above 1,500, resulting in  120,000 for training and 50,525 for validation.

 \vspace{-1em}
\subsection{Input data representation}
Only 1D sequence data are assumed available. 3D structures of proteins, compounds, or their complexes are not used.  

\vspace{-1em}
\subsubsection{Compound data representation}
\vspace{-1em}
\hfill\\\indent {\bf Baseline representation.} A popular compound representation is based on 1D binary substructure fingerprints from PubChem ~\citep{wang2009pubchem}. Mainly, basic substructures of compounds are used as fingerprints by creating binary vectors of 881 dimensions. 

{\bf SMILES representation.} We used SMILES~\citep{weininger1988smiles} \red{that are short ASCII strings} to represent compound chemical structures based on bonds and rings between atoms. \red{64 symbols are used for SMILES strings in our data.  4 more special symbols are introduced for the beginning or the end of a sequence, padding  (to align sequences in the same batch), or not-used ones. Therefore, we defined a compound ``alphabet" of 68 ``letters".} 
Compared to the baseline representation which uses $k$-hot encoding, canonical SMILES strings fully and uniquely determine chemical structures and are yet much more compact.   

\vspace{-1em}
\subsubsection{Protein data representation}
\vspace{-1em}

\hfill\\\indent {\bf Baseline representation.} Previously the most common protein representation for CPI classification was a 1D binary vector whose dimensions correspond to thousands of  (5,523 in~\citep{tian2016boosting}) Pfam domains~\citep{Pfam2014}  (structural units) and 1's are assigned based on $k$-hot encoding~\citep{tabei2013scalable,cheng2016effectively}.  We considered all  types of Pfam entries  (family, domain, motif, repeat, disorder, and coiled coil) for better coverage of structural descriptions, which leads to 16,712 entries  (Pfam 31.0) as features.  
Protein sequences are queried in batches against Pfam using the web server HMMER  (hmmscan)~\citep{finn2015hmmer} with the default gathering threshold.   

{\bf Structural property sequence  (SPS) representation.}  Although 3D structure data of proteins is often a luxury and their prediction remains a challenge without templates, it has been of much progress to predict protein structural properties from  sequences
~\citep{cheng2005scratch,magnan2014sspro,RaptorX-Property}.
We used SSPro/ACCPro~\citep{magnan2014sspro} to predict secondary structure class  ($\alpha$-helix, $\beta$-strand, and coil) and solvent accessibility  (exposed or not) for each residue and group neighboring residues of the same secondary structure class into secondary structure elements  (SSEs).  
The details and the pseudo-code for SSE are in Algorithm 1 (Supplementary Data).

Each SSE is further classified: solvent exposed if at least 30\% of residues are and buried otherwise; polar, non-polar, basic or acidic based on the highest odds  (for each type, occurrence frequency in the SSE is normalized by background frequency seen in all protein sequences to remove the effect from group-size difference); short if length $L \leqslant 7$, medium if $7 < L \leqslant 15$, and long if $L > 15$.  In this way, we defined 4 \red{separate} alphabets of 3, 2, 4 and 3 letters, respectively to characterize SSE category, solvent accessibility, physicochemical characteristics, and length  (Table~\red{S1}) and combined letters from the 4 alphabets in the order above to create 72 ``words''  (4-tuples) to describe SSEs. 
Pseudo-code for the protein representation is shown as Algorithm 2 in Supplementary Data. \red{Considering the} 4 more special symbols introduced \red{similarly for compound SMILES strings, we flattened the 4-tuples and thus defined a protein SPS ``alphabet" of 76 ``letters".}

The SPS representation overcomes drawbacks of Pfam-based baseline representation: it provides higher resolution of sequence and structural details for more challenging regression tasks, more distinguishability among proteins in the same family, and more interpretability on which protein segments  (SSEs here) are responsible for predicted affinity.  All these are achieved with a much smaller alphabet of size 76
, which leads to around 100-times more compact representation of a protein sequence than the baseline. 
In addition, the SPS sequences are much shorter than amino-acid sequences and prevents convergence issues when training RNN and LSTM for sequences longer than 1,000~\citep{IndRNN}.

\vspace{-1em}
\subsection{RNN for unsupervised pre-training}

We encode compound SMILES or protein SPS into representations, first by unsupervised deep learning from abundant unlabeled data. We used a recurrent neural network  (RNN) model, seq2seq~\citep{sutskever2014sequence}, that has seen much success in natural language processing and was recently applied to embedding compound SMILES strings into fingerprints~\citep{xu2017seq2seq}. 
A Seq2seq model is an auto-encoder that consists of two recurrent units known as the encoder and the decoder, respectively  (see the corresponding box in Fig.~\ref{fig:whole_CPI}). 
The encoder maps an input sequence   (SMILES/SPS in our case) to a fixed-dimension vector known as the thought vector. Then the decoder maps the thought vector to the target sequence  (again, SMILES/SPS here). We choose gated recurrent unit  (GRU)~\citep{cho2014properties} as our default seq2seq model and treat the thought vectors as the representations learned from the SMILES/SPS inputs. 
The detailed GRU configuration and advanced variants  (bucketing, bidirectional GRU, and  attention mechanism which provides a way to ``focus'' for encoders) can be found in Sec. 1.4 of Supplementary Data.

Through unsupervised pre-training, the learned representations capture nonlinear joint dependencies among protein residues or compound atoms that are far from each other in sequence. Such ``long-term'' dependencies are very important to CPIs since corresponding residues or atoms can be close in 3D structures and jointly contribute to intermolecular interactions. 


\vspace{-1em}
\subsection{Unified RNN-CNN for supervised learning}

With compound and protein representations learned from the above unsupervised learning, we solve the regression problem of compound-protein affinity prediction using supervised learning. For either proteins or compounds, we append a CNN after the RNN  (encoders and attention models only) that we just trained. The CNN model consists of a one-dimensional  (1D) convolution layer followed by a max-pooling layer. The outputs of the two CNNs  (one for proteins and the other for compounds) are concatenated and fed into two more fully connected layers.  

The entire RNN-CNN pipeline is trained from end to end~\citep{wang2016studying}, with the pre-trained RNNs serving as warm initializations, for improved performance over two-step training. The pre-trained RNN initializations prove to be very important for the non-convex training process~\citep{sutskever2013importance}. In comparison to such a ``unified'' model, we also include the ``separate" RNN-CNN baseline for comparison, in which we fixed the learned RNN part and train CNN on top of its outputs.  

\subsection{\red{Attention mechanisms in unified RNN-CNN}}
We have also introduced \red{three} attention mechanisms to unified RNN-CNN models.  The goal is to both improve predictive performances and  enable model interpretability at the level of ``letters"  (SSEs in proteins and atoms in compounds) \red{and their pairs}.  

\red{1) Separate attention. This default attention mechanism is applied  to the compound and the protein separately so the attention learned on each side is non-specific to a compound-protein pair. However, it has the least parameters among the three mechanisms.}

\red{2) Marginalized attention. To introduce pair-specific attentions, we first use a pairwise ``interaction'' matrix for a pair and then marginalize it based on maximization over rows or columns for separate compound or protein attention models, which is motivated by \cite{lu2016hierarchical}.} 

\red{3) Joint attention. We have developed this novel attention model to fully explain the pairwise interactions between components (compound atoms and protein SSEs). Specifically, we use the same pairwise interaction matrix but learn to represent the pairwise space and consider attentions on pairwise interactions rather than ``interfaces" on each side. Among the three attention mechanisms, joint attention provides the best interpretability albeit with the most parameters.}

\red{These} attention models  (for proteins,  compounds, \red{or their pairs}) are jointly trained with the RNN encoder and the CNN part.  Learned parameters of theirs include attention weights on all ``letters'' for a given string \red{(or those on all letter-pairs for a given string-pair)}.  Compared to that in unsupervised learning, each attention model here outputs a single vector as the input to \red{its corresponding} subsequent 1D-CNN model.

More details \red{on unified RNN-CNN and attention mechanisms} can be found in Sec. 1.5 of Supplementary Data.

\vspace{-2em}
\section{Results and Discussion}
\subsection{Compound and protein representations}

We compared the auto-encoding performances of our vanilla seq2seq model and 4 variants: bucketing, bi-directional GRU  (``fw+bw''), attention mechanism, and attention mechanism with fw+bw, respectively, in Tables S3 and S4  (Supplementary Data).  We used the common assessment metric in language models, perplexity, which is related to the entropy $H$ of modeled probability distribution $P$  ($\mbox{Perp} (P)=2^{H (P)} \geqslant 1$).  First, the vanilla seq2seq model had lower test-set perplexity for compound SMILES than protein SPS  (\red{7.07} versus 41.03), which echoes the fact that, \red{compared to protein SPS strings}, compound SMILES strings are defined in an alphabet of less letters   ($\red{68}$ versus \red{76}) \red{and are of shorter lengths  (100 versus 152)}, thus their RNN models are easier to learn.  Second, bucketing, the most ad-hoc option among all, did not improve the results much.  Third, whereas bi-directional GRUs lowered perplexity by about 2$\sim$3.5 folds \red{and the default attention mechanism did much more} for compounds or proteins, they together achieved the best performances  (perplexity being \red{1.0002} for compound SMILES and 1.001 for protein SPS). 

Therefore, the last seq2seq variant,  bidirectional GRUs with attention mechanism, is regarded the most appropriate one for learning compound/protein representations and adopted thereinafter.

\vspace{-1em}
\subsection{Compound-protein affinity prediction}

\subsubsection{Comparing novel representations to baseline ones}

To assess how useful the learned/encoded protein and compound representations are for predicting compound-protein affinity, we compared the novel and baseline representations in affinity regression using the labeled datasets.  The representations were compared under the same shallow machine learning models --- ridge regression, lasso regression and random forest  (RF). 

\vspace{-2em}
\begin{table}[!htb]
\resizebox{\columnwidth}{!}{
\red{
\begin{tabular}{l|rrr|rrr}
\toprule
 & \multicolumn{3}{c|}{Baseline representations}  &\multicolumn{3}{c}{Novel representations}\\
\cline{2-7}
& Ridge & Lasso & RF &  Ridge & Lasso & RF\\
\hline
 Training & 1.16 (0.60)& 1.16 (0.60)&0.76 (0.86)&1.23 (0.54)&1.22 (0.55)&\textbf{0.63} (0.91)\\
 \hline
 Testing & 1.16 (0.60)&1.16 (0.60)&\textbf{0.91} (0.78)&1.23 (0.54)&1.22 (0.55)&\textbf{0.91} (0.78)\\
   \hline\hline
 ER & 1.43 (0.30)&1.43 (0.30)&1.44 (0.37)&1.46 (0.18)&1.48 (0.18)&\textbf{1.41} (0.26)\\
 \hline
 Ion Channel& 1.32 (0.22)&1.34 (0.20)&1.30 (0.22)&1.26 (0.23)&1.32 (0.17)&\textbf{1.24} (0.30)\\
   \hline
 GPCR & \textbf{1.28} (0.22)&1.30 (0.22)&1.32 (0.28)&1.34 (0.20)&1.37 (0.17)&1.40 (0.25)\\
    \hline
 Tyrosine Kinase & \textbf{1.16} (0.38)&1.16 (0.38)&1.18 (0.42)&1.50 (0.11)&1.51 (0.10)&1.58 (0.11)\\
 \hline\hline
 Time  (core hours) & 3.5& 7.4&1239.8 &0.47&2.78&668.7\\
 \hline
 Memory  (GB) & 7.6& 7.6& 8.3&7.3&7.3&6.3\\
\bottomrule
\end{tabular}
}
}
\caption{Comparing the novel representations to the baseline based on RMSE  (and Pearson correlation \red{coefficient r}) of pIC$_{50}$ shallow regression.}
\label{table:Rep_comp}
\end{table}
\vspace{-2em}
From Table~\ref{table:Rep_comp} we found that our novel representations learned from SMILES/SPS strings by seq2seq models  outperform baseline representations of $k$-hot encoding of molecular/Pfam features. For the best performing random forest models,  using \red{46}\% less training time and \red{24}\% less memory, the novel representations achieved the same performance over the default test set as the baseline ones and lowered root mean squared errors  (RMSE) for two of the \red{four} generalization sets whose target protein classes  (nuclear estrogen receptors / ER and ion channels) are not included in the training set.  
Similar improvements were observed on \red{p$K_i$, p$K_d$, and pEC$_{50}$}  predictions in Tables~\red{S5--7}  (Supplementary Data), respectively. These results show that learning protein and compound representations from even unlabeled datasets alone could improve their context-relevance for various labels.  \red{We also note that, unlike  Pfam-based protein representations that exploit curated information only available to some proteins and their homologs, our SPS representations do not assume such information and can apply to uncharacterized proteins lacking annotated homologs.}

\vspace{-1em}
\subsubsection{Comparing shallow and deep models}

Using the novel representations we next compared the performances of affinity regression between the best shallow model  (random forest) and various deep models.  For both separate and unified RNN-CNN models, we tested results from a single model with (hyper)parameters optimized over the training/validation set, averaging a ``parameter ensemble'' of 10 models derived in the last 10 epochs, and averaging a ``parameter+NN'' ensemble of models with varying number of neurons in the fully connected layers  ((300,100), (400,200) and  (600,300)) trained in the last 10 epochs.  \red{The attention mechanism used here is the default, separate attention.} 


From Table~\ref{table:genaral_acc} we noticed that unified RNN-CNN models outperform both random forest and separate RNN-CNN models  (\red{the} similar performances \red{between RF and separate RNN-CNN} indicated a potential to further improve RNN-CNN models with deeper models).  By using a relatively small amount of labeled data  (which are usually expensive and limited), protein and compound representations learned from abundant unlabeled data can be tuned to be more task-specific.  We also noticed that averaging an ensemble of unified RNN-CNN models further improves the performances especially for some generalization sets of ion channels and GPCRs. As anticipated, averaging ensembles of models reduces the variance  originating from network architecture and parameter optimization thus reduces expected generalization errors.  Similar observations were made for p$K_i$ predictions as well (Table~\red{S8} in Supplementary Data) even when their hyper-parameters were not particularly optimized and simply borrowed from pIC$_{50}$ models.  Impressively, unified RNN-CNN models without very deep architecture could predict IC$_{50}$ values with relative errors below $10^{0.7}$=5 fold  (or 1.0~kcal/mol) for the test set and even \red{around $10^{1.3}=20$ fold  (or 1.8~kcal/mol)} on average for protein classes not seen in the training set.  \red{Interestingly, GPCRs and ion channels had similar RMSE but more different Pearson's $r$, which is further described by the distributions of predicted versus measured pIC$_{50}$ values for various sets (Fig.~S5 in Supplementary Data).}

\begin{table*}[!htb]
\centering
{
\red{
\begin{tabular}{l|l|lll|lll}
\toprule
&  & \multicolumn{3}{c|}{Separate RNN-CNN Models}  &\multicolumn{3}{c}{Unified RNN-CNN Models}\\ \cline{3-8}
  &  RF  & single & parameter &parameter+NN & single & parameter  & parameter+NN\\
    &   &  & ensemble&ensemble&  & ensemble  & ensemble\\
  \hline
 Training & 0.63 (0.91) & 0.68 (0.88)&0.67 (0.90)&0.68 (0.89)&0.47 (0.94)&0.45 (0.95)&\textbf{0.44} (0.95)\\
 \hline
 Testing & 0.91 (0.78) & 0.94 (0.76) & 0.92 (0.77)&0.90 (0.79)&0.78 (0.84)&0.77 (0.84)&\textbf{0.73} (0.86)\\
   \hline
 Generalization -- ER & \textbf{1.41} (0.26)& 1.45 (0.24) & 1.44 (0.26)&1.43 (0.28)&1.53 (0.16)&1.52 (0.19)&1.46 (0.30)\\
 \hline
 Generalization -- Ion Channel & \textbf{1.24} (0.30)& 1.36 (0.18) &1.33 (0.18)&1.29 (0.25)&1.34 (0.17)&1.33 (0.18)&1.30 (0.18)\\
   \hline
 Generalization -- GPCR &  1.40 (0.25)& 1.44 (0.19)&1.41 (0.20)&1.37 (0.23)&1.40 (0.24)&1.40 (0.24)&\textbf{1.36} (0.30)\\
    \hline
 Generalization -- Tyrosine Kinase &  1.58 (0.11) & 1.66 (0.09)&1.62 (0.10)&1.54 (0.12)&1.24 (0.39)&1.25 (0.38)&\textbf{1.23} (0.42)\\
 \hline
\end{tabular}
}
}
\caption{Under novel representations learned from seq2seq, comparing random forest and variants of separate RNN-CNN and unified RNN-CNN models based on RMSE  (and Pearson correlation \red{coefficient $r$}) for pIC$_{50}$ prediction.}
\label{table:genaral_acc}
\end{table*}
\vspace{-1em}

\begin{table*}[!htb]
\centering
\resizebox{2\columnwidth}{!}{
\red{
\begin{tabular}{l|lll|lll|lll}
\toprule
& \multicolumn{3}{c|}{Separate attention}  &\multicolumn{3}{c|}{Marginalized attention} &\multicolumn{3}{c}{Joint attention}\\ \cline{2-10}
   & single & parameter &parameter+NN & single & parameter  & parameter+NN& single & parameter  & parameter+NN\\
    &   & ensemble&ensemble&  & ensemble  & ensemble&  & ensemble  & ensemble\\
  \hline
 Training &0.47 (0.94)&0.45 (0.95)&0.44 (0.95)&0.50 (0.94)&0.47 (0.95)&0.42 (0.96)&0.48 (0.94)&0.44 (0.94)&\textbf{0.40} (0.95)\\
 \hline
 Testing & 0.78 (0.84)&0.77 (0.84)&\textbf{0.73} (0.86)&0.81 (0.83)&0.79 (0.84)&\textbf{0.73} (0.86)&0.84 (0.82)&0.80 (0.83)&\textbf{0.73} (0.86)\\
   \hline
 Generalization -- ER & 1.53 (0.16)&1.52 (0.19)&1.46 (0.30)&1.69 (0.20)&1.67 (0.20)&1.53 (0.30)&1.78 (0.03)&1.68 (0.04)&\textbf{1.37} (0.23)\\
 \hline
 Generalization -- Ion Channel &1.34 (0.17)&1.33 (0.18)&\textbf{1.30} (0.18)&1.63 (0.01)&1.64 (0.06)&1.41 (0.13)&1.54 (0.25)&1.53 (0.26)&1.42 (0.26)\\
   \hline
 Generalization -- GPCR &1.40 (0.24)&1.40 (0.24)&\textbf{1.36} (0.30)&1.59 (0.17)&1.57 (0.18)&1.42 (0.24)&1.53 (0.19)&1.53 (0.19)&1.38 (0.25)\\
    \hline
 Generalization -- Tyrosine Kinase &1.24 (0.39)&1.25 (0.38)&\textbf{1.23} (0.42)&1.69 (0.22)&1.62 (0.25)&1.50 (0.32)&2.22 (0.18)&2.17 (0.21)&2.04 (0.17)\\
 \hline
\end{tabular}
}
}
\caption{\red{Under novel representations learned from seq2seq, comparing different attention mechanisms of unified RNN-CNN models based on RMSE (and Pearson correlation \red{coefficient $r$}) for pIC$_{50}$ prediction.}}
\vspace{-1em}
\label{table:attention}
\end{table*}

\subsubsection{\red{Comparing attention mechanisms in prediction}}
\red{To assess the predictive powers of the three attention mechanisms introduced, we compared their pIC$_{50}$ predictions in Table~\ref{table:attention} using the same dataset and the same unified RNN-CNN models as before.  All attention mechanisms had similar performances on the training and test sets. However, as we anticipated, separate attention with the least parameters edged joint attention in generalization (especially for receptor tyrosine kinases). Meanwhile, joint attention had similar predictive performances and much better interpretability, thus will be further examined in all interpretability studies in case studies for selective drugs.}

\vspace{-1em}
\subsubsection{Deep transfer learning for new classes of protein targets} 

Using the generalization sets, we proceed to explain and address our unified RNN-CNN models' relatively worse performances for new classes of protein targets without any training data. We \red{chose to analyze separate attention models with the best generalization results and} first noticed that proteins in various sets have different distributions in the SPS alphabet (4-tuples). In particular, the test set, \red{ion channels/GPCRs/tyrosine kinases}, and estrogen receptors are increasingly different from the training set  (measured by Jensen-Shannon distances \red{in SPS letter or SPS length distribution}) (Fig.~S3 in Supplementary Data), which correlated with increasingly deteriorating performance relative to the training set  (measured by the relative difference in RMSE) with a Pearson correlation coefficient of \red{0.68 (SPS letter distribution) or 0.96 (SPS length distribution)} (Fig.~S4 in Supplementary Data).  

\begin{figure*}[!htb]
    \centering
\includegraphics[width=1\textwidth]{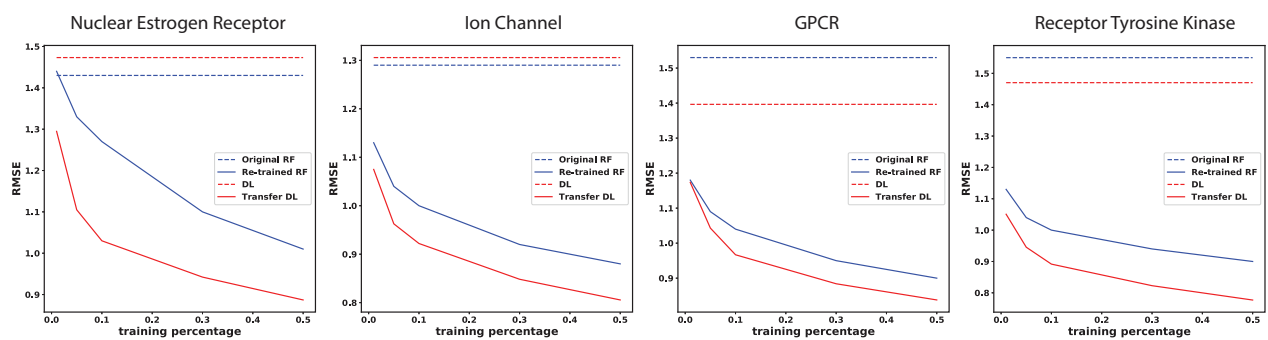}
    \caption{\red{Comparing  strategies to generalize predictions for four sets of new protein classes: original random forest (RF), original param.+NN ensemble of unified RNN-CNN models (DL for deep learning with the default attention), and re-trained RF or transfer DL using incremental amounts of labeled data in each set.}}
    \label{fig:Transfer_learning}
\end{figure*}
 
To improve the performances for new classes of proteins, we compare two strategies: re-training shallow models  (random forest) from scratch based on new training data alone and ``transferring'' original deep models  (unified parameter+NN ensemble \red{with the default separate attention}) to fit new data  (see details in Supplementary Data). The reason is that new classes of targets often have few labeled data that might be adequate for re-training class-specific shallow models from scratch but not for deep models with much more parameters. 


As shown in Fig.~\ref{fig:Transfer_learning}, deep transfer learning models increasingly improved the predictive performance compared to the original deep learning models, when increasing amount of labeled data for new protein classes are made available.  The improvement was significant even with  $1\%$ training coverage for \red{each new protein class.} Notably, deep transfer learning models outperformed random forest models that were re-trained specifically for each new protein class.  

\vspace{-1em}
\subsection{Predicting target selectivity of drugs}
We went on to test how well our unified RNN-CNN models could predict certain drugs' target selectivity, using 3 sets of drug-target interactions of increasing prediction difficulty. Our novel representations and models successfully predicted target selectivity \red{for 6 of 7 drugs} whereas baseline representations and shallow models (random forest) failed for \red{most drugs}.

\vspace{-1em}
\subsubsection{Factor Xa versus Thrombin}

Thrombin and factor X  (Xa) are important proteins in the blood coagulation cascade. Antithrombotics, inhibitors for such proteins, have been developed to treat cardiovascular diseases. Due to thrombin's other significant roles in cellular functions and neurological processes, it is desirable to develop inhibitors specifically for factor Xa.  DX-9065a is such a selective inhibitor  (p$K_i$ value being 7.39 for Xa and $<$2.70 for thrombin)~\citep{brandstetter1996x}. 

\vspace{-2em}
\begin{table}[H]
\centering
\resizebox{1\columnwidth}{!}{
\red{
\begin{tabular}{l|c|c|c|c}
\toprule
& \multicolumn{1}{c|}{Baseline rep. + RF}  &\multicolumn{1}{c|}{Novel rep. + RF} &\multicolumn{1}{c}{Novel rep. + DL (sep. attn.)}  &\multicolumn{1}{|c}{Novel rep. + DL (joint attn.)}\\ \hline
 Thrombin &6.36  &\textbf{6.71}&5.68&4.77\\
 Factor Xa &\textbf{6.87} &6.54&\textbf{8.08}&\textbf{8.64} \\
\bottomrule
\end{tabular}
}
}
\caption{Predicted \red{p$K_i$} values and target specificity for compound DX-9065a interacting with human factor Xa and thrombin.}
\label{thrombin Specificity}
\end {table}
\vspace{-2em}
\red{We used the learned p$K_i$ models in this study. Both proteins (thrombin and factor Xa) were included in the $K_i$ training set with 2,294 and 2,331 samples, respectively, but their interactions with the compound DX-9065a were not.}. Table~\ref{thrombin Specificity} suggested that \red{random forest correctly predicted the target selectivity (albeit with smaller than 0.5-unit difference) using baseline representations but failed to do so using novel representations. In contrast, our models with separate and joint attention mechanisms both correctly predicted the compound's favoring Xa. Moreover, our models predicted selectivity levels being 2.4 (separate attention) and 3.9 (joint attention) in p$K_i$ difference ($\Delta$p$K_i$), where the joint attention model produced predictions very close to the known selectivity margin ($\Delta$p$K_i\geqslant$4.7). }

\vspace{-1em}
\subsubsection{\red{Cyclooxygenase  (COX) protein family}}

\red{COX protein family represents an important class of drug targets for inflammatory diseases.  These enzymes responsible for prostaglandin biosynthesis include  COX-1  and  COX-2  in  human,  both  of  which  can  be  inhibited  by  nonsteroidal  anti-inflammatory drugs  (NSAIDs). We chose three common NSAIDs known for human COX-1/2 selectivity: 
celecoxib  (pIC$_{50}$ for COX-1: 4.09;  COX-2: 5.17),  ibuprofen  (COX-1: 4.92, COX-2: 4.10) and rofecoxib   (COX-1: $<$4; COX-2: 4.6)~\citep{luo2017network}. This is a very challenging case for selectivity prediction because selectivity levels of all NSAIDs are close to or within 1 unit of pIC$_{50}$.} 

\red{We used the learned pIC$_{50}$ ensemble models in this study. COX-1 and COX-2 both exist in our IC$_{50}$ training set with 959 and 2,006 binding examples, respectively, including 2 of the 6 compound-protein pairs (ibuprofen and celecoxib with COX-1 individually).}
\vspace{-2em}
\begin{table}[H]
\centering
\resizebox{\columnwidth}{!}{
\red{
\begin{tabular}{c|ccc|ccc|ccc|ccc}
\toprule
& \multicolumn{3}{c|}{Baseline rep. + RF}  &\multicolumn{3}{c|}{Novel rep. + RF} &\multicolumn{3}{c|}{Novel rep. + DL (sep. attn.)} &\multicolumn{3}{c}{Novel rep. + DL (joint attn.)}\\ \cline{2-13}
 & CEL & IBU & ROF & CEL & IBU & ROF & CEL & IBU & ROF & CEL & IBU & ROF\\
\midrule
 COX-1 & 6.06 & 5.32 & 5.71  &6.41&6.12&6.13&5.11 &\textbf{6.06} &5.67 &5.18 & \textbf{5.94} & 6.00\\
 COX-2 &6.06 &5.32 &5.71 &\textbf{6.57}&\textbf{6.19}&\textbf{6.21}&\textbf{7.60} &5.96 &\textbf{6.51} &\textbf{7.46}&5.62&\textbf{6.03}\\
\bottomrule
\end{tabular}
}
}
\caption{\red{Predicted pIC$_{50}$ values and target specificity for three NSAIDs (CEL: celecoxib, IBU: ibuprofen and ROF: rofecoxib) interacting with human COX-1 and COX-2.}}
\label{COX Specificity}
\end {table}
\vspace{-2em}
\red{From Table~\ref{COX Specificity}, we noticed that, using the baseline representations, random forest incorrectly predicted COX-1 and COX-2 to be equally favorable targets for each drug.  This is because the two proteins are from the same family and their representations in Pfam domains are indistinguishable. Using the novel representations, random forest correctly predicted target selectivity for two of the three drugs  (celecoxib and rofecoxib), whereas our unified RNN-CNN models (both attention mechanisms) did so for all three. Even though the selectivity levels of the NSAIDs are very challenging to predict, our models were able to predict all selectivities correctly with the caveat that few predicted differences might not be statistically significant (for instance, the 0.03-unit difference for rofecoxib using joint attention).}

\subsubsection{Protein-tyrosine phosphatase  (PTP) family}

Protein-tyrosine kinases and protein-tyrosine phosphatases  (PTPs) are controlling reversible tyrosine phosphorylation reactions which are critical for regulating metabolic and  mitogenic signal transduction processes. Selective PTP inhibitors are sought for the treatment of various diseases including cancer, autoimmunity, and diabetes. Compound 1 [2- (oxalyl-amino)-benzoic acid or OBA] and its derivatives, compounds 2 and 3  (PubChem CID: 44359299 and 90765696) 
, are highly selective toward PTP1B rather than other proteins in the family such as PTPRA, PTPRE, PTPRC and SHP1 ~\citep{iversen2000structure}. \red{Specifically, the p$K_i$ values of OBA, compound 2, and compound 3 against PTP1B are 4.63, 4.25, and 6.69, respectively; and their p$K_i$ differences to the closest PTP family protein are 0.75, 0.7, and 2.47, respectively ~\citep{iversen2000structure}.}

\red{We used the learned p$K_i$ ensemble models in this study. PTP1B, PTPRA, PTPRC, PTPRE and SHP1 were included in the $K_i$ training set with 343, 33, 16, 6 and 5 samples respectively. These examples just included OBA binding to all but SHP1 and compound 2 binding to PTPRC.}

\begin{table}[!htb]
\centering
\resizebox{\columnwidth}{!}{
\red{
\begin{tabular}{l|lll|lll|lll|lll}
\toprule
& \multicolumn{3}{c|}{Baseline rep. + RF}  &\multicolumn{3}{c|}{Novel rep. + RF} &\multicolumn{3}{c|}{Novel rep. + DL (sep. attn.)}  &\multicolumn{3}{c}{Novel rep. + DL (joint attn.)}\\
 \cline{2-13}
 Protein& Comp1 & Comp2 & Comp3  & Comp1 & Comp2 & Comp3  & Comp1 & Comp2 & Comp3& Comp1 & Comp2 & Comp3\\
\midrule
 {\bf PTP1B}&4.15 &3.87& 5.17 &6.70&6.55&6.71&3.76&\textbf{3.84}&\textbf{3.92}&2.84&\textbf{4.10}&\textbf{4.04}\\
 PTPRA& 4.15 &3.87& 5.17&6.29&6.59&6.27&2.73&2.90&3.44&2.39&2.62&2.12 \\
 PTPRC& 4.15 &3.87& 5.17&\textbf{6.86}&6.73&\textbf{6.87}&3.37&3.25&3.19&3.36&3.49&2.97\\
 PTPRE& 4.15 &3.87& 5.17&6.79&6.68&6.81&\textbf{3.83}&3.75&3.85&2.75&2.93&2.61\\
 SHP1& 4.15 &3.87& 5.17&6.71&\textbf{6.74}&6.73&3.37&3.38&3.89&\textbf{3.42}&3.52&3.22\\
 \bottomrule
\end{tabular}
}
}
\caption{Predicted \red{p$K_i$} values and target specificity for three PTP1B-selective compounds interacting with five proteins in the human PTP family.}
\label{PTP1B Specificity}
\vspace{-2em}
\end {table}
Results in Table~\ref{PTP1B Specificity} showed that random forest using baseline representations cannot tell \red{binding affinity differences} within the PTP family as the proteins' Pfam descriptions are almost indistinguishable.  Using novel representations, random forest \red{incorrectly predicted target selectivity for all 3 compounds, whereas unified RNN-CNN models with both attention mechanisms correctly did so for all but one (compound 1 -- OBA). We also noticed that, although the separate attention model predicted likely insignificant selectivity levels for compounds 2 ($\Delta$p$K_i=0.09$) and 3 ($\Delta$p$K_i=0.03$), the joint attention model much improved the  prediction of selectivity margins  ($\Delta$p$K_i=0.58$ and 0.82 for compounds 2 and 3, respectively) and their statistical significances. }

\vspace{-1em}
\subsection{Explaining target selectivity of drugs}
After successfully predicting target selectivity for some drugs, we proceed to explain using attention scores how our deep learning models did so and what they reveal about those compound-protein interactions.  

\vspace{-1em}

\subsubsection{How do the compound-protein pairs interact?}

Given that SPS and SMILES strings are interpretable and attention models between RNN encoders and 1D convolution layers can report their focus, we pinpoint SSEs in proteins and atoms in compounds with high attention scores, which are potentially responsible for CPIs.  To assess the idea, we chose 3 compound-protein pairs that have 3D crystal complex structures from the Protein Data Bank; and extracted  residues \red{in direct contacts with ligands} (their SSEs are regarded ground truth for binding site) for each protein from ligplot diagrams provided through PDBsum~\citep{de2013pdbsum}.  
Based on \red{joint attention scores $\alpha_{ij}$'s on pairs of protein SSE $i$ and compound atom $j$} from the single unified RNN-CNN model
, we picked the top 10\% (4) SSEs as predicted binding sites.  \red{Specifically, we first corrected joint attention scores to be $\beta_{ij}=\alpha_{ij}-\left(\sum_{k=1}^{I}\alpha_{kj}\right)/I \quad (\forall i=1,\ldots,I, \quad j=1,\ldots, J)$ to offset the contribution of any compound atom $j$ with promiscuous attentions over all protein SSEs.  We then calculated the  attention score $\beta_{i}$ for protein SSE $i$ by max-marginalization ($\beta_{i}=\max_{j}\beta_{ij}$). No negative $\beta_{i}$ was found in this case thus no further treatment was adopted.} 

\vspace{-1em}

\begin{table}[!htb]
\centering
\resizebox{\columnwidth}{!}{
\red{
\begin{tabular}{c|c|c|c|c|c|c|c}
\toprule
& &\multicolumn{2}{c|}{Number of SSEs}  &\multicolumn{4}{c}{Top 10\% (4) SSEs predicted as binding site by joint attn.}  \\
\cmidrule(l){3-4}\cmidrule(l){5-8}
Target--Drug Pair & PDB ID& total &binding site &  \# of TP & Enrichment & Best rank & P value\\ \hline
Human COX2--rofecoxib & 5KIR&40& 6&1&1.68&4&1.1e-2\\
\hline
Human PTP1B--OBA & 1C85&34& 5&1&1.70&1&1.1e-10\\
\hline
Human factor Xa--DX9065 & 1FAX&31& 4 &3 &5.81&2&2.2e-16\\
\midrule
 \hline
\bottomrule
\end{tabular}
}
}
\caption{Interpreting deep learning models: predicting binding sites based on joint attentions.}
\label{tab:interpret-bind}
\vspace{-1em}
\end {table}

Table~\ref{tab:interpret-bind} shows that, compared to randomly ranking the SSEs, our approach can enrich binding site prediction by \red{1.7}$\sim$\red{5.8} fold for the three CPIs. \red{Consistent with the case of target selectivity prediction, joint attention performed better than separate attention did (Table S9). One-sided paired $t$-tests (see details in Sec. 1.7 of Supplementary Data) suggested that binding sites enjoyed higher attention scores than non-binding sites in a statistically significant way. When the strict definition of binding sites is relaxed to residues within 5\AA\ of any heavy atom of the ligand, results were further improved with all top 10\% SSEs of factor Xa being at the binding site (Table S10).}

We delved into the predictions for factor Xa--DX-9065a interaction in Fig.~\ref{fig:1fax}  (the other 2 are in Fig.~S6 of Supplementary Data). Warmer colors (higher attentions) are clearly focused near the ligand. The red \red{loops connected through a $\beta$ strand (resi. 171--196) were} correctly predicted to be at the binding site with a high rank 2, thus a true positive  (TP).  The SSE ranked first, a false positive, is its immediate neighbor in sequence \red{(resi. 162-170; red helix at the bottom) and is near the ligand. In fact, as mentioned before, when the binding site definition is relaxed, all top 10\% SSEs were at the binding site.  } Therefore, in the current unified RNN-CNN model with attention mechanism, wrong attention could be paid to sequence neighbors of ground truth; and additional information  (for instance, 2D contact maps or 3D structures of proteins, if available) could be used as additional inputs to reduce false negatives. 

\red{We also max-marginalized $\beta_{ij}$ over protein SSE $i$ for $\beta_j$ -- attention score on atom $j$ of the compound. Many high attention scores were observed for compound atoms (Fig.~S7), which is somewhat intuitive as small-molecule compounds usually fit in protein pockets or grooves almost entirely. The top-ranked atom happened to be a nitrogen atom forming a hydrogen bond with an aspartate (Asp189) of factor Xa, although more cases need to be studied more thoroughly for a conclusion.}

\begin{figure}[!htb]
    \vspace{-2em}
    \centering
    \includegraphics[width=0.6\columnwidth]{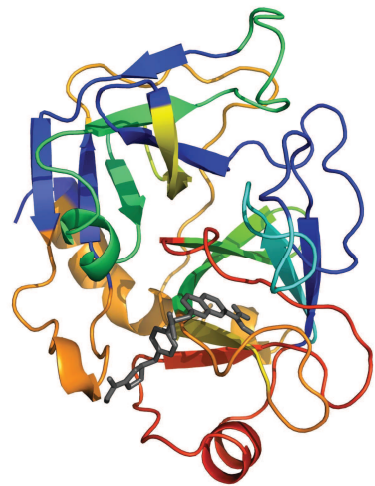}
        \vspace{-2em}
    \caption{\red{Interpreting deep learning models for factor Xa binding-site prediction based on joint attention: 3D structure of factor Xa  (colored cartoons including helices, sheets, and coils) in complex with DX-9065a  (black sticks)  (PDB ID:1FAX) where protein SSEs are color-coded by attention scores  ($\beta_i$) where warmer colors indicate higher attentions.}}
    \label{fig:1fax}
    \vspace{-2em}
\end{figure}

\subsubsection{How are targets selectively interacted?}
To predictively explain the selectivity origin of compounds, we designed an approach to compare attention scores between pairs of CPIs and tested it using factor Xa-selective DX-9065a with known specificity origin. 

For selective compounds that interact with factor Xa over thrombin, position 192 has been identified: it is a charge-neutral polar glutamine  (Gln192) in Xa but a negatively-charged glutamate  (Glu192) in thrombin~\citep{huggins2012rational}. DX-9065a exploited this difference with a carboxylate group forming unfavorable electrostatic repulsion with Glu192 in thrombin but favorable hydrogen bond with Gln192 in Xa.  To compare DX-9065a interacting with the two proteins, we performed amino-acid sequence alignment between the proteins and split two sequences of mis-matched SSEs  (count: 31 and 38) into those of perfectly matched segments  (count: 50 and 50). In the end, segment 42, where SSE 26 of Xa and SSE 31 of thrombin align,  is the ground truth containing position 192 for target selectivity.

\begin{figure}[!htb]
    \centering
    \includegraphics[width=1\columnwidth]{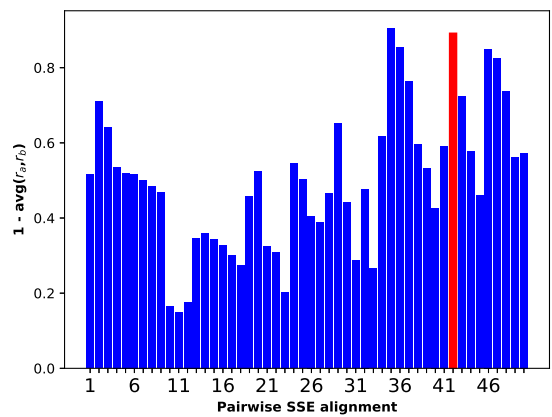}
    \caption{\red{Interpreting deep learning models for factor Xa specificity based on joint attentions.  Pairwise alignment of amino-acid sequences of factor Xa and thrombin decomposed both sequences into 50 segments  (labeled by indices).  These segments are scored by one less the average of the corrected attention rank ratios for the two compound-protein interactions. The ground truth of specificity origin is in red.}} 
    \vspace{-1em}
    \label{fig:factorXa_spec}
\end{figure}

For DX-9065a interacting with either factor Xa or thrombin, we ranked the SSEs based on the attention scores from the unified RNN-CNN single model and assigned each segment the same rank as its parent SSE. \red{Due to the different SSE counts between thrombin and factor Xa, we normalized each rank for segment $i$ by the corresponding SSE count for a rank ratio $r^i$. For each segment we then substracted from 1 the average of rank ratios between factor Xa and thrombin interactions so that highly attended segments in both proteins can be scored higher.} Fig.~\ref{fig:factorXa_spec} shows that the ground-truth segment \red{in red} was ranked the \red{2nd} among 50 segments \red{albeit with narrow margins over the next 3 segments.}

\section{Discussion}
We lastly explore alternative representations of proteins and compounds and discuss remaining challenges. 
\subsection{Protein representations using amino acid sequences}
\red{As shown earlier, our SPS representations integrate both sequence and structure information of proteins and are much more compact compared to the original amino acid sequences. That being said, there is a value to consider a protein sequence representation with the resolution of residues rather than SSEs: potentially higher-resolution precision and interpretability.  We started with unsupervised learning to encode the protein sequence representation with seq2seq. 
More details are given in Sec.~1.8 of Supplementary Data.} 

\vspace{-1em}
\begin{table}[H]
\resizebox{\columnwidth}{!}{
\centering
\red{
\begin{tabular}{c|c|c}
  \hline
  &  SPS rep. +attention+fw/bw& seq. rep. +attention+fw/bw\\
  \hline
 Training error (Perplexity) &\textbf{1.003} & 11.46\\
 \hline
 Testing error (Perplexity) &\textbf{1.001} & 12.69\\
 \hline
 Time  (h) & \textbf{96} & 192\\
 \hline
\end{tabular}
}
}
\caption{\red{Comparing the auto-encoding performances between amino acid and SPS sequences using the best seq2seq model (bidirectional GRUs with attention mechanism).}}
\label{table:seq_rep}
\vspace{-2em}
\end{table}

\red{Compared to SPS representations, protein sequences are 10-times longer and demanded 10-times more GRUs in seq2seq, which suggests much more expensive training. Under the limited computational budget, we trained the protein sequence seq2seq models using   twice the time limit on the SPS ones. The perplexity for the test set turned out to be over 12, which is much worse than 1.001 in the SPS case (see Sec.~3.1) and deemed inadequate for subsequent (semi-)supervised learning. Learning very long sequences is challenging in general and calls for advanced architectures of sequence models.}

\subsection{Unified RNN/GCNN-CNN for protein SPS strings and compound graphs}
\red{We have chosen SMILES representations for compounds partly due to recent advancements of sequence models especially in the field of natural language processing. Meanwhile, the descriptive power of SMILES strings can have limitations.  For instance, some syntactically invalid SMILES strings can still correspond to valid chemical structures.  Therefore, we also explore chemical formulae (2D graphs) for compound representation.}  

\red{We replaced RNN layers for compound sequences with graph CNN (GCNN) in our unified model (separate attention) and kept the rest of the architecture. This new architecture is named unified RNN/GCNN-CNN.  The GCNN part is adopting a very recently-developed method  \citep{gao2018interpretable} for compound-protein interactions.  More details can be found in Sec.~1.9 of Supplementary Data.  
}

\vspace{-1em}
\begin{table}[H]
\centering
\resizebox{\columnwidth}{!}{
\red{
\begin{tabular}{l|lll|lll}
\toprule
& \multicolumn{3}{c}{SMILES rep.}  &\multicolumn{3}{c}{Graph rep.} \\ \cline{2-7}
   & single & parameter &parameter+NN & single & parameter  & parameter+NN\\
    &   & ensemble&ensemble&  & ensemble  & ensemble\\
  \hline
 Training &0.47 (0.94)&0.45 (0.95)&\textbf{0.44} (0.95)&0.55 (0.92)&0.54 (0.92)&0.55 (0.92)\\
 \hline
 Testing & 0.78 (0.84)&0.77 (0.84)&\textbf{0.73} (0.86)&1.50 (0.35)&1.50 (0.35)&1.34 (0.45)\\
   \hline
 Generalization -- ER & 1.53 (0.16)&1.52 (0.19)&\textbf{1.46} (0.30)&1.68 (0.05)&1.67 (0.03)&1.67 (0.07)\\
 \hline
 Generalization -- Ion Channel &1.34 (0.17)&1.33 (0.18)&\textbf{1.30} (0.18)&1.43 (0.10)&1.41 (0.13)&1.35 (0.12)\\
   \hline
 Generalization -- GPCR &1.40 (0.24)&1.40 (0.24)&\textbf{1.36} (0.30)&1.63 (0.04)&1.61 (0.04)&1.49 (0.07)\\
    \hline
 Generalization -- Tyrosine Kinase &1.24 (0.39)&1.25 (0.38)&\textbf{1.23} (0.42)&1.74 (0.01)&1.71 (0.03)&1.70 (0.03)\\
 \hline
\end{tabular}
}
}
\caption{\red{Comparing unified RNN-CNN (SMILES strings for compound representation) and unified RNN/GCNN-CNN (graphs for compound representation)  based on RMSE (and Pearson's correlation coefficient) for pIC$_{50}$ prediction.}}
\vspace{-2em}
\label{table:graph}
\end{table}

\red{Results in Table~\ref{table:graph} indicate that the unified RNN/GCNN-CNN model using compound graphs did not  outperform the unified RNN-CNN model using compound SMILES in RMSE and did a lot worse in Pearson's correlation coefficient. These results did not show the superiority of SMILES versus graphs for compound representations \textit{per se}.  Rather, they show that graph models need new architectures and further developments to address the challenge. We note recent advancements in deep graph models~\citep{GilmerSRVD17,Coley17,VAE-graph18}.}

\section{Conclusion}

We have developed accurate and interpretable deep learning models for predicting compound-protein affinity using only compound identities and protein sequences.  By taking advantage of massive unlabeled compound and protein data besides labeled data in semi-supervised learning, we have jointly trained unified RNN-CNN models for learning context- and task-specific protein/compound representations and predicting compound-protein affinity.  These models outperform baseline machine-learning models.  And impressively, they achieve the relative error of IC$_{50}$ within 5-fold for a comprehensive test set and even that within 10-fold for generalization sets of protein classes unknown to the training set. Deeper models would further improve the results. Moreover, for the generalization sets, we have devised transfer-learning strategies to significantly improve model performance using as few as 40 labeled samples.  

Compared to conventional compound or protein representations using molecular descriptors or Pfam domains, the encoded representations learned from novel structurally-annotated SPS sequences and SMILES strings improve both predictive power and training efficiency for various  machine learning models. Given the novel representations with better interpretability, we have included attention mechanism in the unified RNN-CNN models to quantify how much each part of proteins or compounds are focused while the models are making the specific  prediction for each compound-protein pair. 

When applied to case studies on drugs of known target-selectivity, our models have successfully predicted target selectivity in all cases whereas conventional compound/protein representations and machine learning models have failed some. Furthermore, our analyses on attention weights have shown promising results for predicting protein binding sites as well as the origins of binding selectivity, thus calling for further method development for better interpretability.  

\red{For protein representation,} we have chosen SSE as the resolution for interpretability due to the known sequence-size limitation of RNN models~\citep{IndRNN}. One can easily increase the resolution to residue-level by simply feeding to our models amino-acid sequences  (preferentially of length below 1,000) instead of SPS sequences, \red{but needs to be aware of the much increased computational burden and much worse convergence when training RNNs.} \red{For compound representation, we have started with 1D SMILES strings and have also explored 2D graph representations using graph CNN (GCNN).  Although the resulting unified RNN/GCNN-CNN model did not improve against unified RNN-CNN, graphs are more descriptive for compounds and more developments in graph models are needed to address remaining challenges. }

\section*{Acknowledgments}
This project is in part supported by the National Institute of General Medical Sciences of the National Institutes of Health  (R35GM124952 to YS) and the Defense Advanced Research Projects Agency  (FA8750-18-2-0027 to ZW). Part of the computing time is provided by the Texas A\&M High Performance Research Computing.   

\vspace{-2em}

\bibliographystyle{natbib}

\bibliography{Ref}

\end{document}